\documentclass{article}
\usepackage{authblk} 

\usepackage[authoryear,round,longnamesfirst]{natbib}
\title{How COVID-19 has Impacted American Attitudes Toward China: A Study on Twitter}

\author[1]{Gavin Cook}
\author[1]{Junming Huang}
\author[1]{Yu Xie\thanks{Corresponding author: yuxie@princeton.edu. The authors thank Susan T. Fiske, Xuechunzi Bai, and Gandalf Nicolas Ferreira at Princeton University for valuable suggestions on the Stereotype Content Model. We thank Phillip Rush, Karla M. Perez-Gazca, Rebecca J. Han, Lucas J. Irwin, Hamna Khurram, Louison M. Sall, Ryan Sung, Christian Venturella, and Sandoval E. Wood at Princeton University for coding tweets.}}
\affil[1]{Paul and Marcia Wythes Center on Contemporary China and Department of Sociology, Princeton University}

\usepackage{amsmath}
\usepackage{graphicx}
\usepackage{rotating}

\usepackage{pdflscape}

\usepackage{lipsum}

\usepackage{multirow}
\usepackage[utf8]{inputenc}
\usepackage{xcolor}
\newcommand{\beginsupplement}{%
        \setcounter{table}{0} 
        \renewcommand{\thetable}{S\arabic{table}} 
        \setcounter{figure}{0}
        \renewcommand{\thefigure}{S\arabic{figure}} 
     }

\begin{document}
\maketitle

\begin{abstract}
Past research has studied social determinants of attitudes toward foreign countries. Confounded by potential endogeneity biases due to unobserved factors or reverse causality, the causal impact of these factors on public opinion is usually difficult to establish. Using social media data, we leverage the suddenness of the COVID-19 pandemic to examine whether a major global event has causally changed American views of another country. We collate a database of more than 297 million posts on the social media platform Twitter about China or COVID-19 up to June 2020, and we treat tweeting about COVID-19 as a proxy for individual awareness of COVID-19. Using regression discontinuity and difference-in-difference estimation, we find that awareness of COVID-19 causes a sharp rise in anti-China attitudes. Our work has implications for understanding how self-interest affects policy preference and how Americans view migrant communities.
\end{abstract}

\section{Introduction}
The average American does not know a whole lot about the rest of the world \citep{holsti_public_2004}. Americans harbor latent, inchoate images about foreign peoples and places that are informed by their own values and other heuristics \citep{powlick_defining_1998} and only activated in the presence of external primes or triggers \citep{converse_nature_2006}. While leaders in other countries may assume that Americans closely follow the intrigues of their home countries, this is not the case \citep{anholt_place_2010}. Many Americans simply gloss over news about foreign countries or foreign policies due to a lack of interest \citep{graber_processing_1988, gans_deciding_2004}.

Regardless of how under- or mis-informed its citizenry may be, the US remains a democracy, and the views that ordinary Americans hold toward foreign peoples and places can have strong effects on American foreign policy \citep{baum_war_2015}. This is because public demands affect the policy decisions of elected officials \citep{shapiro_public_2011, page_effects_1983}. This is true generally and, more specifically, in the realm of foreign policy. The opinions of business elites may have a stronger impact on foreign policy outcomes than the opinions of the general public, but the attitudes of the general public still have a measurable and significant influence \citep{jacobs_who_2005}. Examples abound of presidents or officials taking less-than-optimal action to appease constituents back home, such as Bill Clinton supposedly advocating for the expansion of NATO to appease voters of Central and Eastern European descent \citep{alison_mitchell_clinton_1996}. Public opinion may play a role in maintaining peace between democracies \citep{tomz_public_2013} and may be more important now than during the Cold War, when the Soviet Union was as a unifying enemy in the American mind \citep{holsti_public_2004}. The views of the American people on China today are important because they constrain and shape the possible actions and stances that the American state may adopt when engaging with China in the future. This is especially significant in an international environment fraught with the potential for armed conflict \citep{allison_destined_2018}.

We seek to understand how the 2020 outbreak of the COVID-19 pandemic in the US may have impacted American views on China by analyzing a large corpus of data on the posts of users of the social media platform Twitter. The unique exogeneity and magnitude of COVID-19's effects on the US provide a unique opportunity to draw on causal inference methods and answer outstanding questions in the literature on self-interest and foreign policy. We use regression discontinuity and difference-in-difference methods to estimate a causal effect of awareness of COVID-19 on sentiment toward China, controlling for time of exposure and geographic location. We use tweeting about COVID-19 as a proxy for awareness of COVID-19 and find that awareness of COVID-19 leads to an increase in anti-China sentiment. 

\subsection{Two reactions to COVID-19}
We hypothesize that Americans reacted to the 2020 coronavirus outbreak in two distinct phases. Firstly, when it was initially reported to have spread from Wuhan in China, COVID-19 stoked anti-China sentiment by triggering latent and evolved mechanisms for pathogen avoidance. Secondly, as the virus began to spread in the US and ruined the livelihoods of many Americans it spurred a second and more intense wave of anti-China sentiment. Because Americans associated China with COVID-19, they blamed China for the loss of life as well as the damage to their economic well-being and degradation of their lifestyles. Their already negative opinions of China soured further as a result.

\subsection{``Yellow Perils'' and the behavioral immune system}
Humans have both a biological and ``behavioral'' immune system \citep{schaller_behavioral_2011}. The biological immune system fights pathogens once they enter the body, but the behavioral immune system conditions us to keep our distance from sickness by provoking disgust to induce avoidance. Some of these disgust reactions may have evolutionary roots, but the behavioral immune system can induce humans to discriminate against outgroups intensely. For evidence of this, we see that a higher disease burden from communicable illness is associated with elevated rates of outgroup hostility and avoidance in the US ~\citep{oshea_infectious_2020}. 

Not only does disease seem to drive discrimination in general, but humans also tend to associate specific diseases with specific outgroups. Syphilis was called the ``French disease'' by the Germans, the ``German disease'' by the Polish, the ``Polish disease'' by the Russians, and the ``Christian disease'' by the Ottoman-era Turks \citep{tampa_brief_2014}. The Chinese people have similarly been associated with plagues and pestilence throughout history. Racially and biologically charged epithets like ``Yellow Peril'' have been used to refer to people of Sinitic descent in the US ~\citep{tchen_yellow_2014, frayling_yellow_2014} and abroad~\citep{bille_introduction_2018}. Following COVID-19's entry into the US, Americans of Chinese descent and phenotypically-similar but otherwise unrelated Americans of Northeast Asian descent were associated with COVID-19 and consequently suffered from elevated rates of both discrimination and hate-motivated crime \citep{sills_preventing_2020, gover_anti-asian_2020}. This strongly suggests that Americans associated East Asians with COVID-19 and were disgusted as a result.

\subsection{Self-interest and foreign policy}
COVID-19 started as a foreign issue but quickly became a domestic crisis. The virus began to spread rapidly from March 2020 onward in the US. Within the first four months of the pandemic, 100,000 Americans had lost their lives to the pandemic \citep{dave_sanders_four_2020}. This number rose to 500,000 by February 2021, one year after COVID-19 first arrived on American shores \citep{lucy_tompkins_entering_nodate}. 

As of June 7, 2021, 34 million Americans had been infected with the disease, and 612,000 Americans had perished as a result of their infection \citep{101073pnas2014746118_united_2021}. The disease disproportionately affects communities of color, and it is estimated that it reduces the lifespans of members of the Black and Latinx communities by three to four times more than it does for the White population \citep{andrasfay_reductions_2021}. Further consequences include impairment of mental health, including elevated suicide rates \citep{sher_impact_2020} and increased economic anxiety \citep{mann_personal_2020}, and the persistent neurological symptoms experienced by sufferers of ``long COVID,'' many of which manifest as ``brain fog'' and reduced mental acuity \citep{graham_persistent_2021}. 

While the loss of life and degradation of health caused by COVID-19 is tragic beyond measure, Americans also suffered devastating economic consequences as a direct result of the pandemic. Consumer spending fell in all states \citep{dong_personal_2021}. Stay-at-home orders caused a 50\% reduction in trips to non-essential businesses and a 19\% drop in revenue for small businesses \citep{alexander_stay-at-home_2020}. This drop was more pronounced in higher-income ZIP codes, in which small business revenue dropped by 65\% \citep{chetty_economic_2020}. By April 2020, employment rates had dropped by 21\% relative to pre-pandemic levels, and this drop disproportionately affected small businesses and low-wage workers \citep{cajner_us_2020}. Overall, the GDP of the US nosedived from 21.5 trillion to 19.5 trillion dollars between the first and second quarter of 2020, representing a staggering loss of more than 2 trillion dollars \citep{us_bureau_of_economic_analysis_table_nodate}. The pandemic caused widespread economic devastation that impacted the pocketbooks and, at minimum, lifestyles of almost every American. 

As we will demonstrate later in the paper, there was a secondary wave of negative attitudes toward China when COVID-19 began to affect American lives. Many Americans associated COVID-19 with China in the first months of COVID lockdowns. There were conspiracy theories expounding its origin in and its spread from China. Donald Trump, then American president, insisted firmly on calling COVID-19 the ``China Virus'’ despite the media and much of the populace calling him a xenophobe. As children were pulled from schools and customers kept from businesses, many Americans blamed China for their misfortunes. We argue that calculations of self-interest informed American images of foreign policy and drove many Americans to dislike China. 
 
An active literature has been devoted to understanding the relationship between self-interest and political behavior, with two main camps adopting opposing views. On one side of the debate are scholars who argue that self-interest does not strongly influence political behavior. These scholars argue that voters behave with ``sociotropic,'' loosely meaning societal- or national-level, interests in mind \citep{meehl_selfish_1977, kinder_sociotropic_1981}. Findings in this vein include \citet{citrin_public_1997}, who show that an individual’s economic situation does not predict his or her views on immigration restrictionism but that his or her appraisal of the nation’s economic situation does \citep{sears_limited_1990}. Work showing that self-interest does matter, however, is much more common. \citet{erikson_caught_2011} demonstrate that American men drafted to fight in the Vietnam War developed an understandable opposition to war and, as a consequence, became more liberal and voted more consistently for the Democratic Party later in life. \citet{wolpert_self-interest_1998} similarly find that self-interest motivates gun owners to oppose bans on firearms. \citet{bobo_opposition_1993} argue that calculations of group-based self-interest lead Whites to oppose race-based redistributive measures. \citet{campbell_self-interest_2002} finds that wealthier seniors are more politically active than other senior citizens but, because they rely less on social security than their less-fortunate peers, do not vote with social security in mind and are not mobilized to vote by appeals to social security as an issue. This extends to farmers, who participate in politics more eagerly than their socioeconomic status would otherwise predict \citep{lewis-beck_agrarian_1977}, and people who lost their jobs during the Great Recession, who are more likely to support welfare programs after experiencing economic hardship \citep{margalit_explaining_2013}. 

It is possible to reconcile these conflicting claims with reference to psychological models of political behavior. \citet{chong_when_2001} show that priming subjects with vignettes or questions that invoke respondents to reflect on their own economic situation leads to greater influence of self-interest on voting behavior. Some voters, moreover, may find it difficult to figure out the influence a particular policy might have on their daily lives. Many voters ``report no connection between government actions and personal experience'' \citep{funk_dual_2000}. This disconnect may lead to behaviors that appear altruistic but are actually motivated by a lack of understanding and, consequently, a lack of self-interest. This is to say that some voters do not think with self-interest in mind because they do not understand how abstract policies impact their own lives. More-educated voters are additionally more able to see the connection between their own self-interest and a given policy proposal and to vote with their self-interest in mind \citep{young_when_1991}. Uninformed voters, then, do not use self-interest to make voting decisions but instead default to value-based heuristics like party affiliation \citep{popkin_reasoning_1994, achen_democracy_2017}. In summary, the impact of a given policy on an individual's personal finances might be so small or so difficult to conclusively discern that that individual may instead default to value-based decision-making.
 
The shock of COVID-19 has been so unprecedentedly large that it elides the distinctions between the sociotropic and self-interest models of political behavior. It matters not if the Twitter users in our data weight individual-level interest more heavily than national-level interest in their decision-making calculus, or vice versa; COVID-19 has negatively affected countries and individual Twitter users alike. Regardless of how a Twitter user defines self-interest, his or her self-interest will have been impacted negatively by COVID-19. Because the impact of COVID-19 was so sudden, widespread, and deeply felt, we are able to use methods of causal inference on our corpus of twitter posts (``tweets''). 


\subsection{Identifying the attitudes of Twitter users toward China}

We draw upon Twitter as a barometer for how the American public views China. Social media data have been used for similar purposes by a number of other scholars. \citet{osmundsen_partisan_2021} conduct a sentiment analysis of American tweets and find that negatively polarized tweeting behavior is predictive of sharing ``fake news.'' \citet{pan_how_2020} analyze tweets of Saudi dissidents to examine how tweets about the Saudi regime changed before and after dissidents were arrested. Twitter data have been used to track how far-right European politicians use fake bot accounts to gain inflated followings \citep{silva_fake_2021} and how the news media influence public discussion of given issues \citep{king_how_2017-1}. Twitter has been used extensively in health and medicine research \citep{sinnenberg_twitter_2017, eichstaedt_psychological_2015}. Posts from Chinese social microblogging sites very similar to Twitter, such as Weibo, have been used to track censorship in mainland China \citep{king_how_2013, king_how_2017}. 
 
We search for tweets that mention China and merge the resulting data with a public dataset of tweets that mention COVID-19 \citep{Chen_2020}. We then identify users whose public profiles reveal locations in the US. Together, the corpus consists of 297 million tweets between January 2017 and June 2020. While we are aware that this might not be a representative sample of all Twitter users \citep{sloan_who_2015}, our results apply to the online discursive environment of the US. We employed eight human coders to manually label $5,000$ tweets with sentiment scores ranging from ``most unfavorable'' to ``most favorable'' (Table~\ref{tab:sentiment-sample}). We then used the tagged tweets to fine-tune BERT, a deep learning model for natural language processing, to automatically identify the sentiment scores of all of the tweets in our corpus. Those scores are presented in a range from -100 (most unfavorable) to 100 (most favorable). See the Data and Methods section for details.

To avoid confusing attitudes toward China with attitudes toward COVID-19, we only measure attitudes toward China on tweets that mention China but not COVID-19. To avoid biasing our attitude measurements toward highly active users, we calculate sentiment scores at the user level before we calculate averages between individuals. We report this two-stage average as the average daily sentiment toward China in the US.
See the Data and Methods section for details.
 
\section{Results}

\subsection{Declining sentiment with increasing tweet volume on China} 

\begin{figure*}
\centering
\caption{\textbf{American attitudes toward China declined to a 3-year low of -38 after the outbreak of COVID-19.} \normalfont{Attitude toward China is measured by averaging sentiment over Twitter users on tweets that mention China but not COVID-19 on a 7-day sliding window. Sudden and steep declines in China-related sentiment occur nationwide. The declines are accompanied by increases in the volume of tweets on China denoted in the figure with light and dark blue shaded areas. The daily number of users mentioning China on Twitter increased from 5,096 in late December 2019 to $66,535$ by the middle of March 2020. On March 16, over $95\%$ of Twitter users who mentioned China also mentioned COVID-19. See Figure~\ref{SI:102-overall-sentiment-volume} for a full version of this image.}}
\label{fig:102-overall-sentiment-volume}
\includegraphics[width=\linewidth]{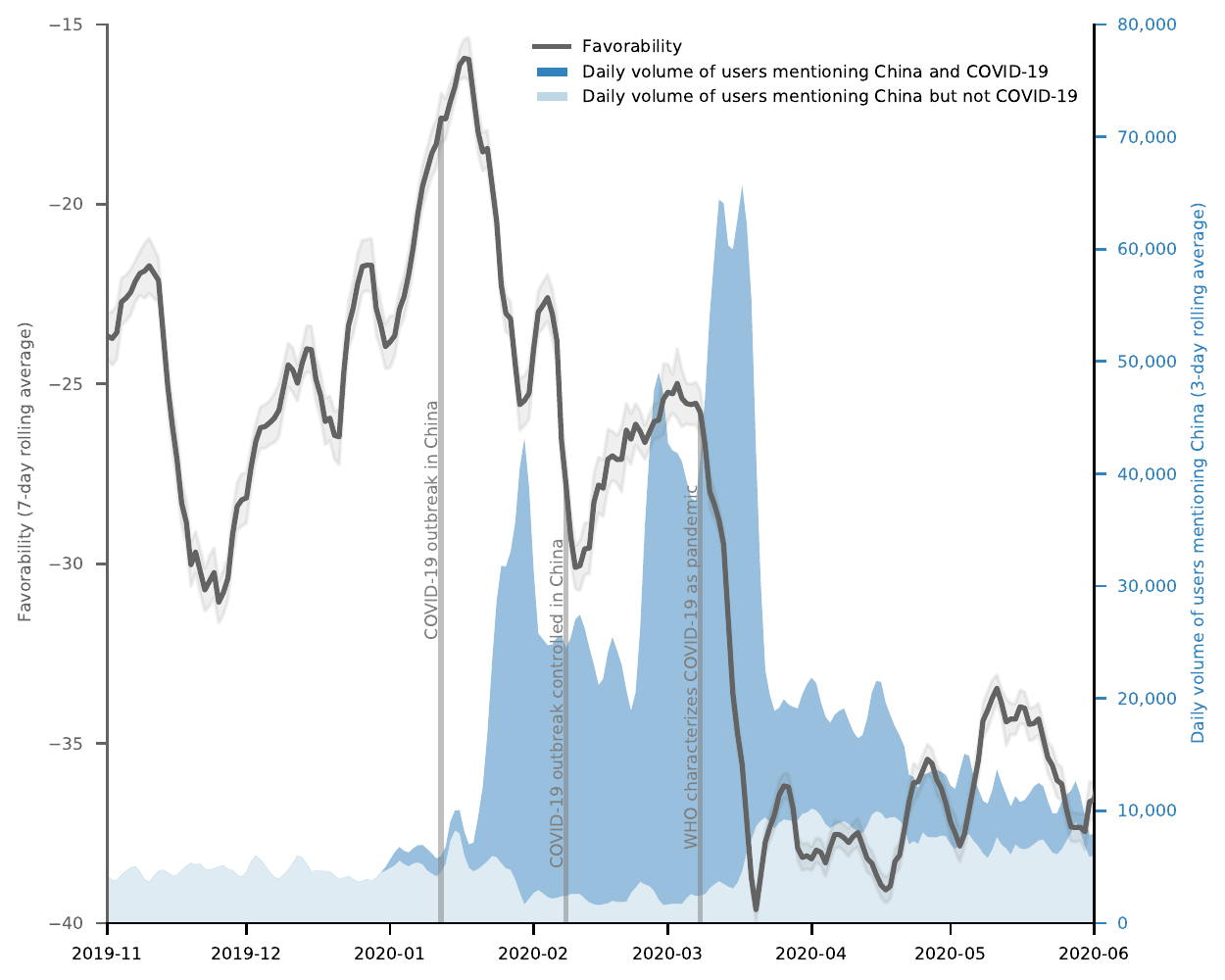}
\end{figure*}

Inchoate attitudes are activated by external triggers, and COVID-19 is a significant trigger for attitudes toward China. China was very much on the minds of the American people in the beginning stages of the pandemic. A virus linked to China was responsible for closing businesses, plucking children from school, and forcing families into their homes. Many of these Americans spoke their minds on Twitter, and tweet volume on China reached a three-year high. Figure~\ref{fig:102-overall-sentiment-volume} displays the sentiment and volume of tweets on China. The solid line (main axis) reveals a two-stage reaction to COVID-19 with two distinct dips in sentiment on China after the outbreak of COVID-19. The secondary axis, denoted in the figure by shades, shows that the number of users tweeting on China increased by tenfold, shooting from less than $6,000$ users per day to a peak of greater than $60,000$ users per day.

The first dip in China-related sentiment in Figure~\ref{fig:102-overall-sentiment-volume} is likely reflective of a disgust reaction provoked by the behavioral immune system. The attitudes of American Twitter users toward China declined sharply within two months after the January 2020 outbreak of COVID-19 in China. This decline was much more intense than any similar dips in China-related sentiment in the past three years covered in our data. 
The average sentiment of China-related tweets, which we quantify in a range from -100 (most unfavorable) to 100 (most favorable), declined from $-15.6$ on January 20 to $-31.7$ on February 17 before eventually reaching a three-year low of $-41.1$ on March 18. Though the reaction of highly individualistic Americans to intense and draconian lockdown policies may have played a part, the rise in anti-Asian hate crimes taken together with the dip in China-related sentiment on Twitter suggests that the outbreak of COVID-19 in Wuhan prompted some Americans to feel negatively toward China and, by an unfortunate conflation, the Chinese people.





\begin{figure}[ht]
   \centering
   \includegraphics{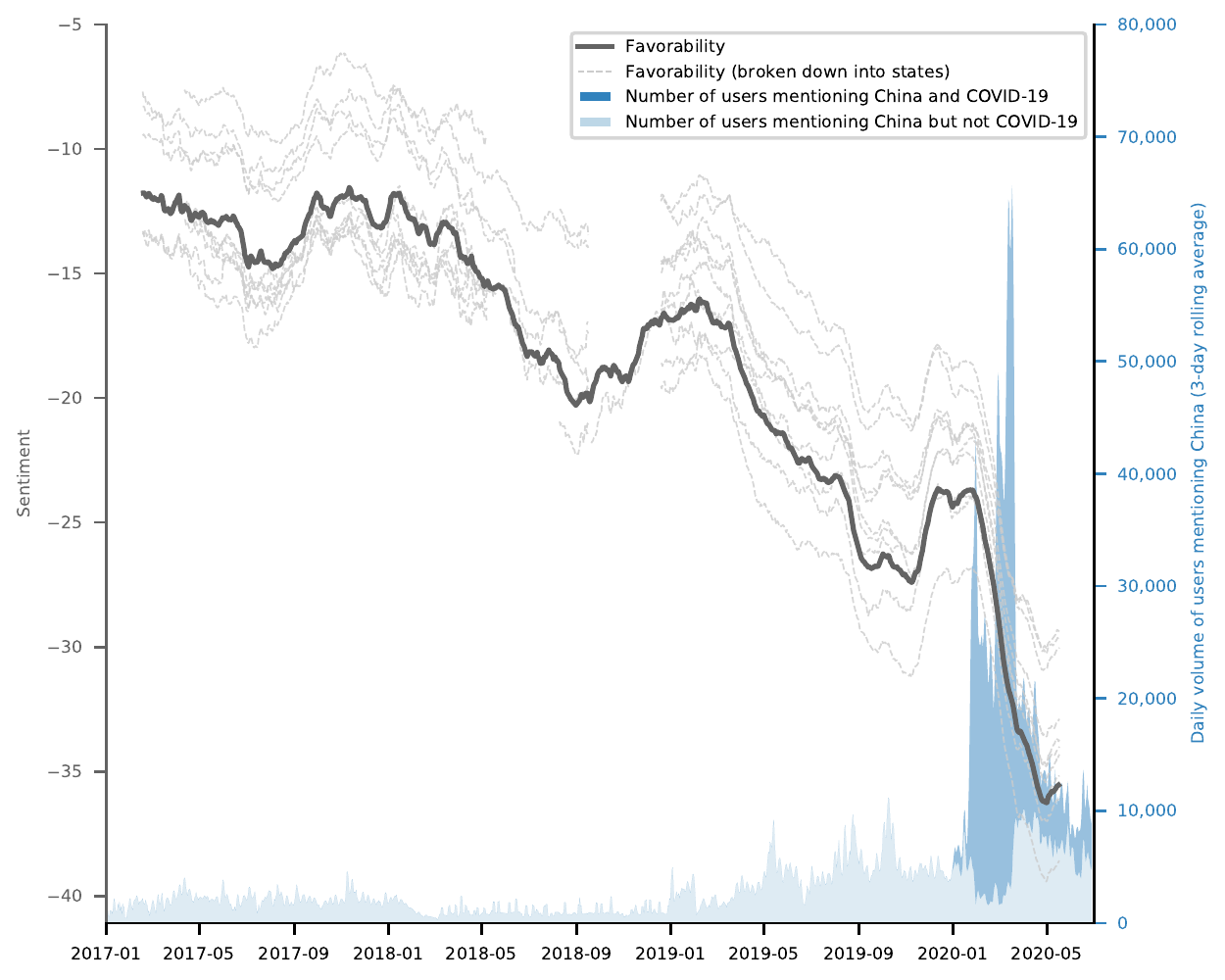}
   \caption{\textbf{A full version of American attitudes toward China from January 2017 to June 2020.} \normalfont{Attitude toward China is measured as an average over Twitter users on tweets mentioning China but not COVID-19 on a 60-day sliding smooth window. It declines slowly before nosediving suddenly after the outbreak of COVID-19 (solid). The decline is accompanied by an increase in tweet volume on China (light and dark blue areas). Dashed lines represent state-level average attitude toward China in the 10 states with the most Twitter users: California, New York, Texas, Florida, District of Columbia, Kansas, Illinois, Washington, Georgia, and Massachusetts.}}
   \label{SI:102-overall-sentiment-volume}
\end{figure}

The sharp drop in China-related sentiment on Twitter after the outbreak of COVID-19 was both sudden and unprecedentedly drastic. In the three years before January 2020, the average weekly sentiment toward China was between $-30$ and $-10$ (Figure~\ref{SI:102-overall-sentiment-volume}). The drop in sentiment toward China that occurred in Jan 2020 was irregular and unprecedentedly sharp. This implies that the increasing prevalence of anti-China attitudes on Twitter is not the result of normal swings but is instead a response to an external trigger.


We also see this decline at the state level. Across the ten states with the most Twitter users, namely California, New York, Texas, Florida, District of Columbia, Kansas, Illinois, Washington, Georgia, and Massachusetts, attitudes toward China declined sharply at almost the same time and to almost the same depth as at the national level (dash lines in Figure~\ref{SI:102-overall-sentiment-volume}). This consistency across states further suggests that the decline in China-related sentiment was not the outcome of random fluctuations but was instead caused by a single external trigger.

\begin{figure*}
\centering
\caption{\textbf{Changes in the composition of Twitter users who mention China.} \normalfont{Users grouped by sentiment toward China before the COVID-19 outbreak (October - December 2019) are displayed on the left side of the figure. $70.56\%$ of users had negative attitudes, $19.44\%$ had neutral attitudes, and $10.00\%$ had positive attitudes. User distribution over sentiment groups after the COVID-19 outbreak (March - May 2020) are on the right side of the figure. $76.91\%$ users were in the negative group, $15.20\%$ were in the neutral group, and $7.90\%$ were in the positive group. A total of $18.38\%$ of users moved to a more negative group (e.g., neutral to negative) while $11.49\%$ moved to a more positive group. $2.18\%$ users moved from the negative group to the positive group, and $3.74\%$ users moved from the positive group to the negative group.}}
\label{fig:304-post-treatment-effect}
\includegraphics[width=\linewidth]{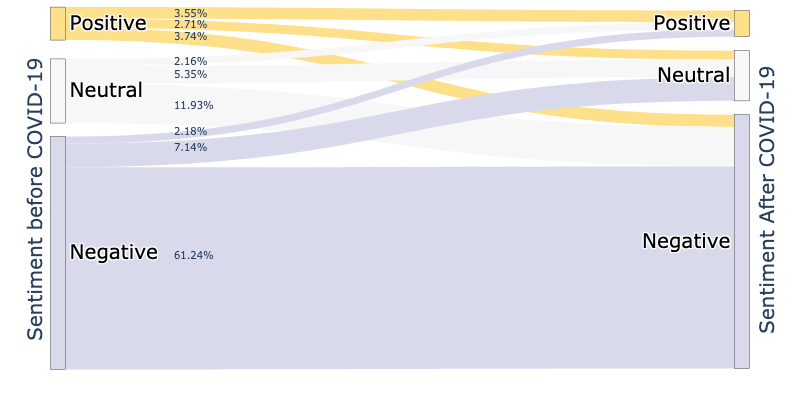}
\end{figure*}

The decline reflects the changing composition of user sentiment as more users adopted negative attitudes toward China. We segment all users into three categories: negative (sentiment from $-100$ to $-20$), neutral (sentiment between $-20$ and $20$), and positive (sentiment from $20$ to $100$) based on their 3-month average attitudes toward China before the outbreak in January 2020. Following the outbreak, $18\%$ of users changed their attitudes toward China to that of a more negative group while only $11\%$ moved to a more positive group. Users in the negative group increased from $71\%$ before the outbreak to $77\%$ after the outbreak while users in the positive group declined from $10\%$ to $8\%$. The number of users who moved from the neutral group to the negative group is almost five times the number of users who moved from the neutral to the positive group  (Figure~\ref{fig:304-post-treatment-effect}).

The sharp decline in sentiment on China was accompanied by a rapid uptick in the number of tweets about China. For three years before January 2020, between $2,000$ to $13,000$ Twitter users located in the US tweeted or retweeted tweets about China per day. As rumors about COVID-19 began to spread in early January 2020, the number of users who mentioned China in tweets began to increase dramatically. Within four weeks, the daily volume of those users reached a peak of $46,792$, over six times of the usual level. After a brief respite toward the end of February 2020, this volume then reached a second peak with a total of $66,535$ users mentioning China on a single day in mid-March, when COVID-19 hit American shores in earnest. These marked increases in tweets about China may be mostly attributed to the outbreak of COVID-19, during which both China and COVID-19 were simultaneously mentioned by nearly $64,000$ users per day in mid-March (dark blue area in Figure~\ref{fig:102-overall-sentiment-volume}). By contrast, the number of users discussing non-COVID China topics stayed almost constant and even decreased in the beginning stage of the pandemic. The imbalanced composition of users tweeting about both COVID-19 and China and those tweeting China but not COVID-19 further implicates COVID-19 in both the increasing volume of tweets and the decline in sentiment toward China.

\subsection{Identifying causality and quantifying treatment effects}
The coincidence of the outbreak of COVID-19 and the rise of anti-China sentiment on social media is remarkable. The exogeneity of COVID-19 allows us to employ two methods to investigate whether COVID-19 is a direct cause of the increase in anti-China attitudes. We examine two hypotheses: a \emph{causal hypothesis} that attributes the rise in anti-China attitudes to the outbreak of COVID-19 and a \emph{non-causal hypothesis} that posits a mere association between the two. We use regression discontinuity and difference-in-difference estimation to adjudicate between these hypotheses. For every Twitter user in our data, each of the two strategies analyzes whether COVID-19 changed their attitude toward China. We find that awareness of COVID-19 causes a rise in anti-China sentiment.


Capitalizing on the timestamps in the Twitter data, we use a \textbf{regression discontinuity} design to model the immediate change in the sentiment of an individual's tweets on China after they were ``exposed'' to information about COVID-19.~\footnote{In the scope of this paper, we use the terms ``treat,'' ``treatment,'' ``control,'' and ``expose'' in the context of causal inference. Although this paper is on COVID-19, we stress that these terms are not intended to refer to medicine in any way and that our paper has nothing to do with the medical reality of COVID-19.}

If there is no causal effect of exposure to information about COVID-19 on expressed sentiment toward China, we would assume that an individual's sentiment toward China should be stable within a small window before and after their exposure to news COVID-19. If, on the other hand, there is a causal effect of knowledge about COVID-19 on views toward China, we would expect to see a discontinuity in an individual's sentiment on China before and after exposure. This assumption is reasonable given that both anti-China sentiment and the volume of tweets on COVID-19 appear to rise gradually when viewed at the week level but spike dramatically when viewed at the year level. This means that a discontinuity in a given user's sentiment toward China would stand out as a strong signal in the noise of gradual weekly change. 

\begin{figure}[ht]
   \centering
   \includegraphics[width=1.0\linewidth]{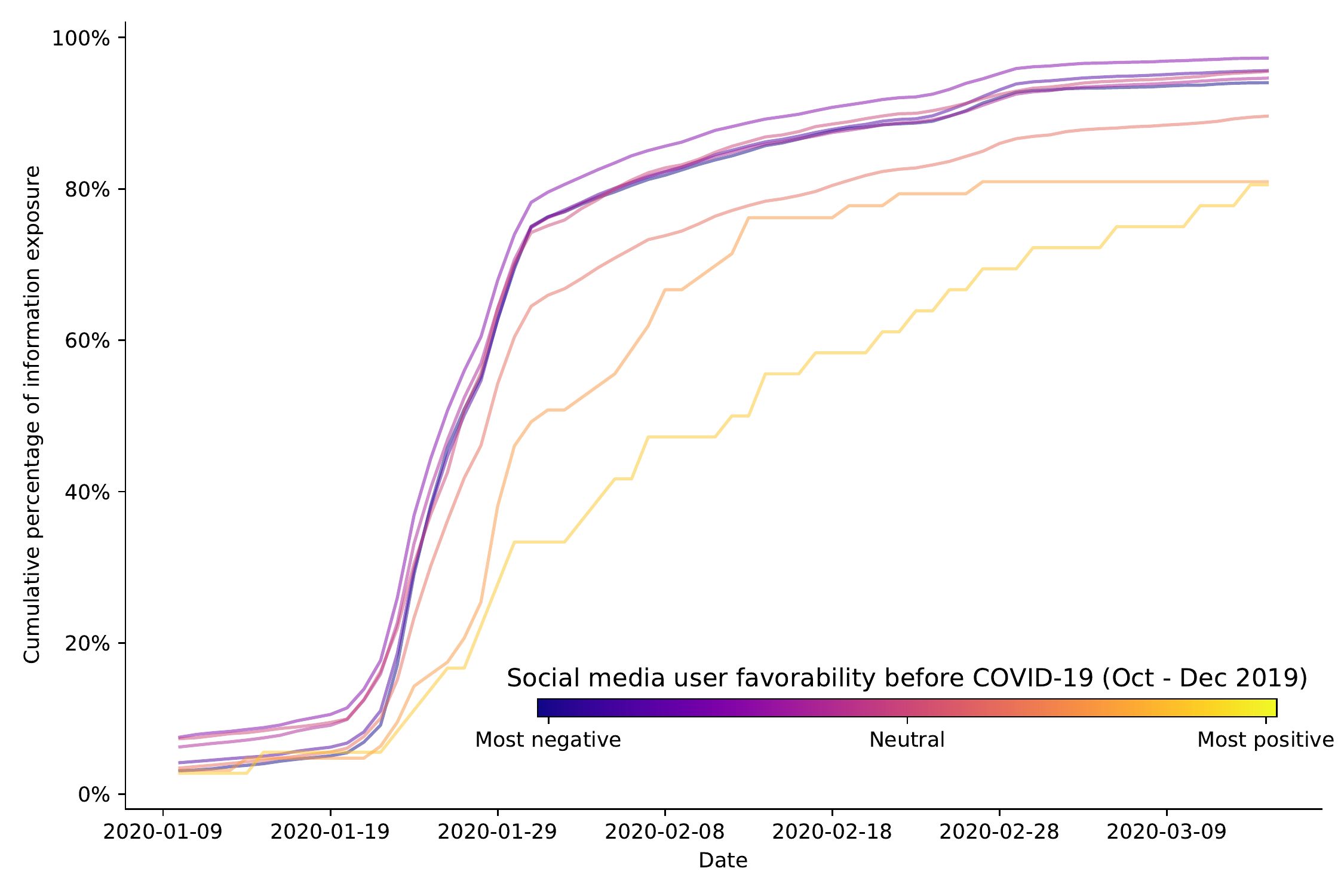}
   \caption{\textbf{Pre-treatment difference.} \normalfont{Twitter users are gradually treated from late January to early March. Users that dislike China (purple) before the outbreak of COVID-19 resopnd to the pandemic early and more quickly compared to Twitter users who favored China before the outbreak (yellow). The y-axis reports the cumulative percentage of Twitter users who have mentioned both China and COVID-19. Anti-China users start to mention COVID-19 in late January, with over 97\% of these users mentioning the virus at least once before the end of February. Meanwhile, pro-China users started to mention the pandemic a bit later, with 90\% users treated by March.}} 
   \label{fig:303-pre-treatment-effect}
\end{figure}

Information about COVID-19 spread to most Twitter users within the initial six weeks of the outbreak from mid-January to late February, and sentiment on China declined steadily over the same time frame. The outbreak eventually garnered the attention of $90\%$ of Twitter users by late February (Figure~\ref{fig:303-pre-treatment-effect}). Though there is a patch of abrupt change in late January, the curves of decline in sentiment and the cumulative percentage of treated users are both relatively smooth. 

We identify the specific day that a Twitter user first posts a COVID-related tweet and compare their sentiment toward China from before and after the day of exposure. This day is taken to be the \emph{treatment} day, and we use it as a proxy for the day the user is first exposed to information about COVID-19. In a typical regression discontinuity design, a researcher would model the individual's sentiment as a temporal continuous function in a neighborhood of the treatment day. This approach, however, necessitates parametric assumptions that require data far away from the treatment threshold. We instead use a conservative non-parametric approach that compares a user's average sentiment toward China one week before the treatment day to that user's sentiment toward China one week after the treatment day. It measures the treatment effect $\Delta_{RD}$ as
$$
\Delta_{RD} = E_d[y_d(t=d) - y_d(t=d-1)],
$$
where $y_d(t)$ denotes the mean attitude in week $t$ averaged over all individuals who are treated in week $d$ and $E_d[\cdot]$ denotes the average over all weeks $d$. At the individual level, the treatment effect is the difference between sentiment pre- and post-treatment, i.e., $y_d(d)- y_d(d-1)$. If the non-causal hypothesis is true, the rise in anti-China sentiment would happen regardless of the presence or absence of information about COVID-19, and the treatment day would not provide any significant information for inferring the tweet sentiment of a specific user. If this were the case, we would predict that a user's attitude toward China would be statistically stable on the days before and after treatment. If the causal hypothesis holds, we would expect to observe a marked discontinuity, specifically a sharp dip, in an individual's attitude toward China after the treatment.
%
%



\begin{figure*}
\centering
\caption{\textbf{Identifying causality with regression discontinuity.} \normalfont{The average sentiment of China-related tweets before and after the date a Twitter user first posts a tweet mentioning COVID-19. Individuals are segmented by treatment week, denoted in vertical gray lines.}}
\label{fig:341c-regression-discontinuity}
\includegraphics[width=\linewidth]{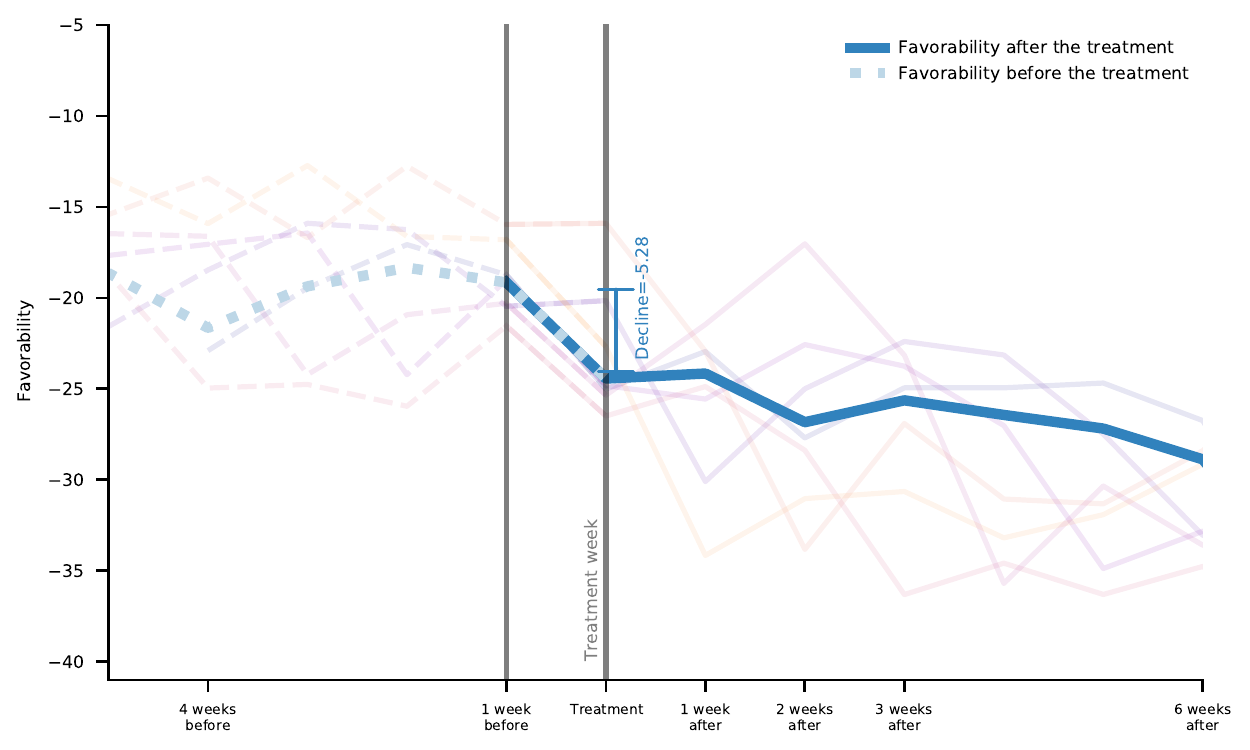}
\end{figure*}

Our empirical results support the causal hypothesis. Figure~\ref{fig:341c-regression-discontinuity} shows that, on average, individuals maintain relatively stable and neutral attitudes toward China before treatment but that their attitudes toward China dip abruptly by $-5.28$ within one week of treatment. Their attitudes then continue to slowly decline for another six weeks before reaching a saturation point. The step-function-like curve reveals a sharp discontinuity around the treatment day, which supports the causal hypothesis. 

\begin{figure}[ht]
   \centering
   \includegraphics{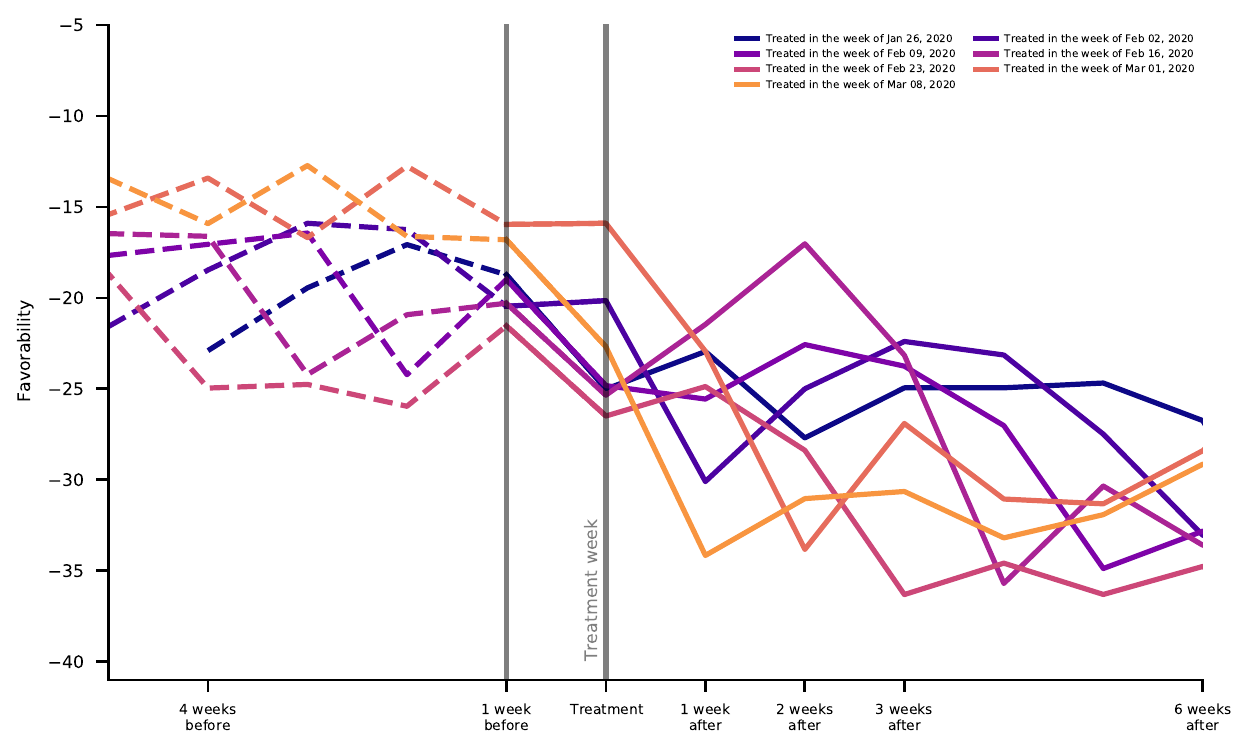}
   \caption{Estimating the treatment effect of tweeting about COVID-19 on sentiment toward China with regression discontinuity. The average sentiment of China-related tweets before and after a Twitter user first posts a tweet mentioning COVID-19.} 
   \label{SI:341c-regression-discontinuity}
\end{figure}

We then group individual users by treatment week and report the change in sentiment for each group (vertical gray lines in Figure~\ref{fig:341c-regression-discontinuity}, full version in Figure~\ref{SI:341c-regression-discontinuity}). The sentiment trends for all groups are broadly similar and consist of a stable level before treatment, a sudden and sharp decline immediately after treatment, and a slow and continued decline to a saturation point. The consistency of this pattern between the groups strongly suggests that timing does not impact the treatment effect. Regardless of when an individual is exposed to information about COVID-19, they experience a similar dip in China-related sentiment and eventually settle at a relatively negative opinion toward China. Because individual-level attitudes toward China are stable before treatment, our empirical results validate the assumption that the day a user first posts about COVID-19 is a robust proxy for treatment.

The regression discontinuity design we used earlier to identify the causal effect of COVID-19 on attitudes toward China is basically, in the language of~\citet{campbell1963experimental}, a ``one-group, pretest-posttest design.'' As such, it lacks a comparison group and thus could suffer from a bias attributable to ``history'' – concomitant events that overlap with treatment. To guard against such biases, we supplement the above regression discontinuity analyses with a difference-in-difference strategy.

The difference-in-difference (DID) strategy compares the change in attitudes toward China over time for the treated group and the change in attitudes over the same time for the control (i.e. untreated) group. Because COVID-19 was and still is a very hot topic on social media, almost $90\%$ of Twitter users had mentioned COVID-19 at least once by March 2020 (Figure~\ref{fig:303-pre-treatment-effect}). If we were to define the control group as users who had not ever mentioned COVID-19, we would be left with a control group so vanishingly small that it would be difficult to use any techniques of statistical inference to compare it to the treated group. We therefore create a DID comparison with moving windows and treat those who had not yet mentioned COVID by the end of each window as eligible controls. For every week in January and February 2020, we create a treated group with all individuals who have mentioned COVID-19 before a given week and a control group with all individuals who have not mentioned COVID-19 by the same week. The overall treatment effect $\Delta_{DID}$ is then measured as the macro-level mean over all windows:
$$
\Delta_{DID} = E_d \left[ \left( y_d(t=d) - y_d(t=d-1) \right) - \left( y_{d^\prime > d}(t=d) - y_{d^\prime > d}(t=d-1) \right) \right],
$$
where $y_d(t)$ measures the macro-level average attitude in week $t$ among individuals who are treated in week $d$, and $y_{d^\prime > d}(t)$ represents the average attitude in week $t$ among individuals who are treated in any week $d^\prime$ after week $d$. In every moving window $d$, we compute $y_d(t=d) - y_d(t=d-1)$, the change in attitudes over time for the treated group, and $y_{d^\prime > d}(t=d) - y_{d^\prime > d}(t=d-1)$, the change in attitudes over time for the control group. The difference in difference estimator $\Delta_{DID}$ should tell us whether knowledge of COVID-19 triggers a Twitter user to tweet negatively about China compared to a counterfactual situation in which he or she had no knowledge of COVID-19.
If the non-causal hypothesis is true, the treated group and the control group would be statistically indistinguishable, and we would predict an expectation of zero for the difference in difference estimator, i.e., $\Delta_{DID} = 0$. If the causal hypothesis holds, the drop in the treated group's attitudes toward China would be significant relative to the ``natural'' change in the control group's attitudes toward China. That is, we would we predict $\Delta_{DID}$ to be non-zero.

\begin{figure*}
\centering
\caption{\textbf{Identifying causality with difference in difference estimation}.\normalfont{The sentiment of Twitter users toward China declines suddenly in the treatment week, i.e. the week after they post their first tweets about COVID-19 (solid lines). In contrast, the counterfactual group, who had not yet been treated by a given treatment week, maintain stable sentiment toward China (dotted lines). The treatment effect is estimated by first taking the difference in sentiment on China before and after the treatment within groups and then taking the difference of these within-group differences between groups. This suggests that tweeting about COVID-19 causes an immediate decline of $4.21$ in sentiment toward China.}}
\label{fig:341b-difference-in-difference}
\includegraphics[width=\linewidth]{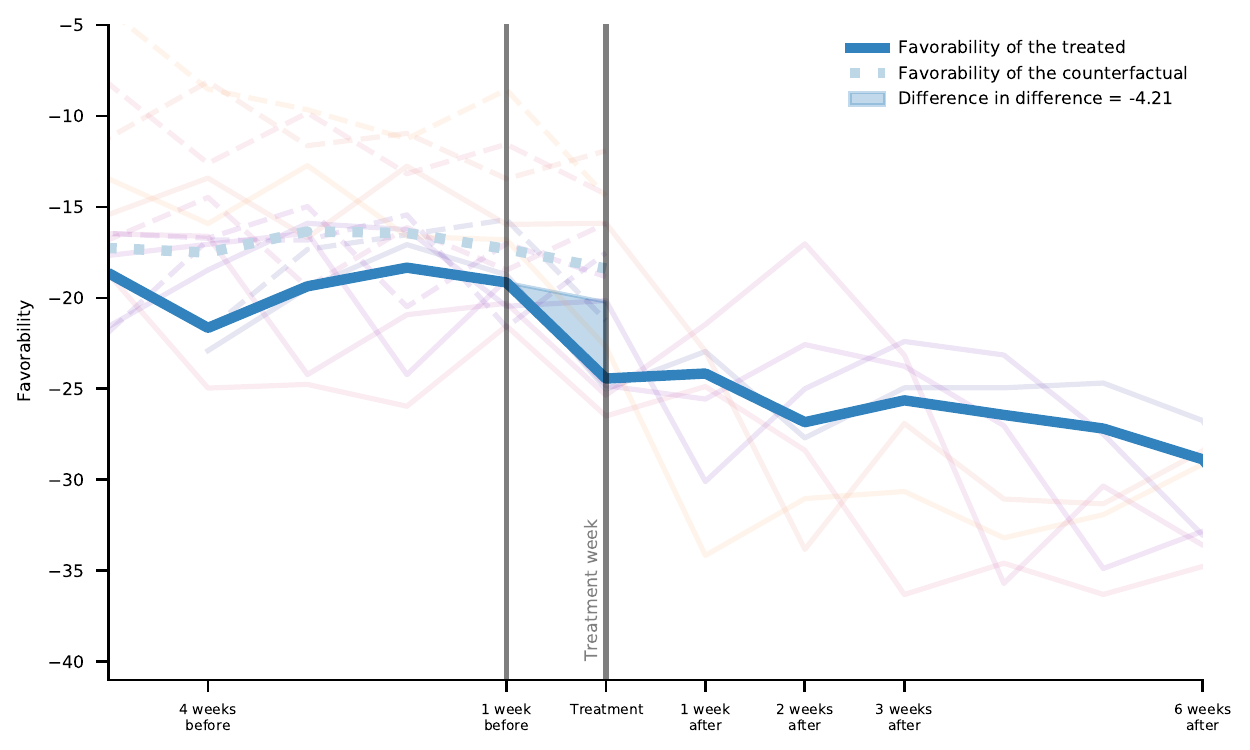}
\end{figure*}

\begin{figure}[ht]
   \centering
   \includegraphics{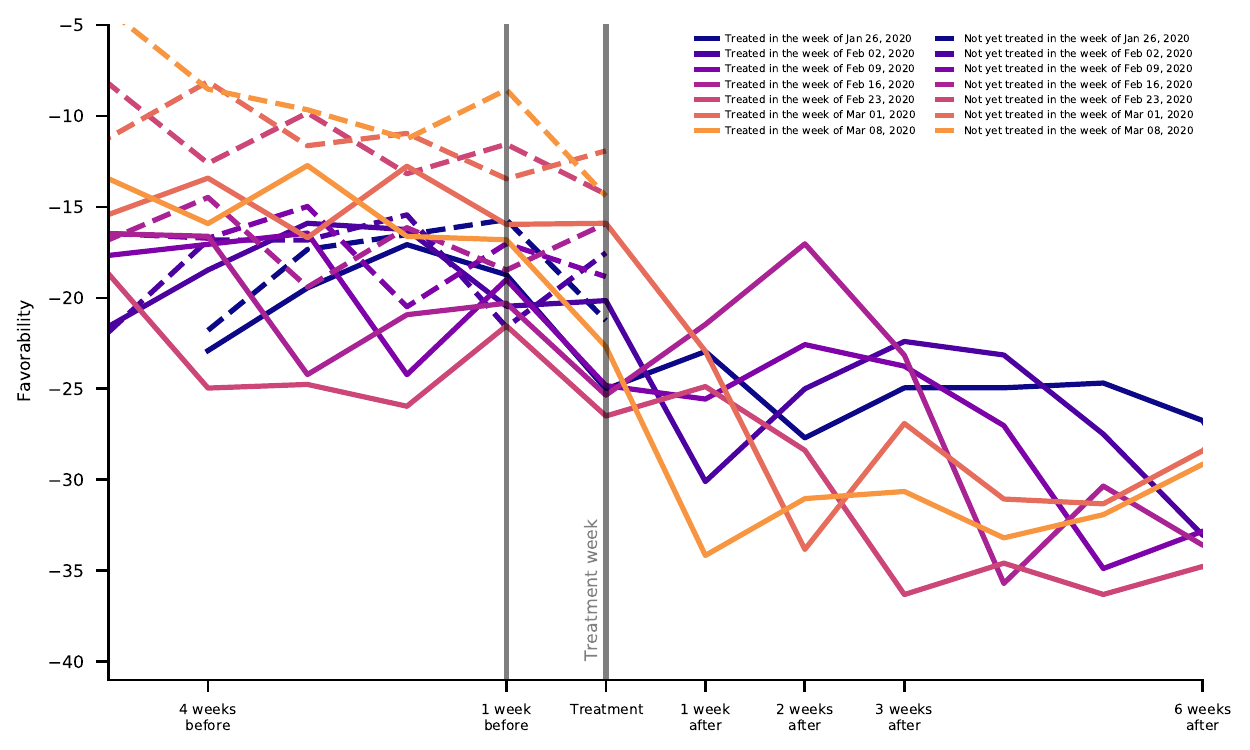}
   \caption{Estimating the treatment effect of tweeting about COVID-19 on attitude toward China with difference in difference estimation. \normalfont{The sentiment of Twitter users on China declines suddenly in the week they first tweet about COVID-19 (solid lines). In contrast, the counterfactual group who had not yet been treated by a given treatment week maintain almost the same level of sentiment toward China (dotted lines). The treatment effect is estimated by first taking the difference in sentiment on China before and after the treatment within groups and then taking the difference of these within-group differences between groups. This estimation strategy suggests that knowledge of COVID-19 causes an immediate decline of $4.21$ in sentiment on China. Furthermore, the effect reaches saturation quickly, as the sentiment of the treated remains unchanged afterwards.}}  
   \label{SI:341b-difference-in-difference}
\end{figure}

Our empirical DID results again support the causal hypothesis. As shown in Figure~\ref{fig:341b-difference-in-difference}, we see an average decline of $\Delta_{DID} = -4.21$ in attitudes toward China in the week following a Twitter user's treatment. The treated group and control group have very similar attitudes toward China until the treatment week, upon which the treated group's attitudes toward China degrade suddenly while the control group's attitudes toward China stay almost constant. A more detailed analysis (Figure~\ref{SI:341b-difference-in-difference}) shows that this pattern is consistent across all of the moving windows we construct. 

\section{Conclusion}
In the wake of the outbreak of COVID-19, American attitudes toward China sunk to a three-year low while tweet volume on China soared to a three-year high. Based on a corpus of over 297 million tweets on China and/or COVID-19, we identify the causal impact of the COVID-19 outbreak on anti-China sentiment. Using regression discontinuity and DID estimation methods, we discover that awareness of COVID-19, measured by proxy as tweeting about COVID-19, leads to a sudden and sharp decline in American attitudes on China. Moreover, we find that the relationship between this awareness and this decline is causal. 

The COVID-19 pandemic and the unique data we have collated offer powerful methodological and theoretical insight. The suddenness of the COVID-19 pandemic is an almost uniquely exogenous trigger for attitudinal change on foreign countries without precedent in the social media era. This creates the opportunity for the assumptions of causal inference to hold in situations where they may have previously been untenable. COVID-19 as a natural experiment specifically allows us to use causal inference methods on a large corpus of social media data. Our analysis reveals that awareness of COVID-19 leads to a marked rise in anti-China sentiment on Twitter. 

Secondly, COVID-19 offers a rare situation where foreign affairs are no longer foreign: the material interests and physical safety of Americans are perceived to be threatened from abroad. Similar incidents are scarce in the American historical record. Apart from the Pearl Harbor and the 9/11 attacks, Americans remain safely isolated behind the natural walls of two oceans and far from military dramas of other continents. Excepting the 1973 oil embargo, the economy of the US has generally been insulated from what happens outside the borders of the US. The COVID-19 pandemic exposed many Americans to a historic and unprecedented combination of economic and physical hardship that they could blame on a foreign source. Substantively, this allows us to test various theories of foreign relations and foreign policy. Following previous work on self-interest and how Americans view foreign countries, we find that Americans did not think much about China before the pandemic, but COVID-19 forced China into the American mind. The COVID-19 pandemic shocked the US in two ways. First, COVID-19 prompted a reaction from the collective behavioral immune system of the American mind that resulted in a wave of disgust toward China. This is reflected in our data by a small but marked peak in the daily number of tweets on China and a worsening in sentiment toward China. Secondly, the livelihoods, lives, and lifestyles of all Americans were impacted by the pandemic as the pandemic left many Americans ill and jobless. This caused a sharp spike in anti-China prejudice. This finding confirms a link between self-interest and political behavior in the realm of foreign policy that has previously been very difficult to test empirically. Given that the American public now views China more negatively than ever before, the American state may adopt anti-China rhetoric and policy positions in the near future.

%
%

\section{Data and methods}\label{section:data-methods}

\textbf{Social media data} are collected from Twitter public messages and profiles to analyze American attitudes toward China. Starting from an open library of near 100 million COVID-19 related tweets~\citep{Chen_2020}, we use an open-source tool to search tweets containing China-related and COVID-19 related keywords. 
Specifically, we search for tweets containing any mention of the words ``China,'' ``Chinese,'' ``Wuhan,'' and ``Beijing'' from January 2017 to June 2020
and tweets containing any mention of the words ``covid,'' ``covid-19,'' ``covid19,'' ``ncov,'' ``ncov19,'' ``coronavirus,'' ``corona+virus,'' ``quarantine,'' ``lockdown,'' ``epidemic,'' ``pandemic,'' ``pandemia,'' ``outbreak,'' ``socialdistancing,'' ``mask,'' ``confirm+case,'' ``death+case,'' ``flatten+curve'' from January 2020 to June 2020.
After merging those two datasets and removing duplicate tweets, we end up with $297,232,539$ unique tweets, $120,250,229$ of which mention China-related words. We further obtain a subset of $68,979,579$ unique English-language tweets mentioning China but not COVID-19. Each tweet bears a time stamp and a unique user identifier, and we anonymized all user identifiers after downloading.
Our dataset contains tweets from 14 million Twitter users, 9 million of which have mentioned China at least once and 8 million of which have mentioned COVID-19 at least once. We search Twitter users who have voluntarily revealed geographic information on their profiles and identified those located in the US. ~\footnote{We recognize that many Twitter users located in the US may not be American citizens, but for the purposes of this paper, we will describe all of the users in this paper as ``Americans.''}
We end up with 174,187 users who (1) are living in the US and (2) have tweeted at least one tweet about China per month from January to April 2020. We analyze individual-level trends of sentiment toward China on those users.


\textbf{Tweet sentiments} are quantitatively inferred with BERT, an open-source state-of-the-art natural language processing model of deep neural networks. We first employed eight research assistants to manually label the expressed China-related sentiments of $5,000$ English-language tweets. These tweets were labeled with an integer-valued favorability score from -2 to 2, i.e., -2 (most unfavorable), -1 (somewhat unfavorable), 0 (neutral), 1 (somewhat favorable), 2 (most favorable). Table~\ref{tab:sentiment-sample} shows examples of those tweets. We then used the labeled tweets to fine-tune a pretrained BERT model to learn a mapping from tweet text to integer favorability score at an accuracy of $95.3\%$. This fine-tuned model assigns favorability scores to all English tweets in the entire corpus. Then we proportionally map the scores from $[-2, 2]$ to $[-100, 100]$ for easier presentation, i.e., -100 (most unfavorable), -50 (somewhat unfavorable), 0 (neutral), 50 (somewhat favorable), 100 (most favorable).

\textbf{Attitudes toward China} are measured on $68,979,579$ tweets mentioning China but not COVID-19. This is because a majority of tweets mentioning China in early 2020 also mention COVID-19, expressing a mixture of attitudes that may be directed at China, COVID-19, or both. In fact, 5 in 6 of the Twitter users who talk about China in February 2020 also mention COVID-19 in the same tweet, and most tweets about COVID-19 are unsurprisingly negative. Given that the rapid increase in overwhelmingly negative tweets on COVID-19 would negatively bias our measurements of China-related sentiment, it is inaccurate to include said tweets when measuring attitudes toward China. We simply ignore all tweets mentioning both China and COVID-19 and report the sentiment toward China in tweets that mention China but not COVID-19 unless otherwise specified.

\textbf{Population attitudes} are calculated as an average sentiment over all individuals.
The posting frequency of social media users is extremely heterogeneous. In our sample, $82\%$ of the users post $40\%$ of the tweets. Averaging sentiment over all tweets would therefore be very biased toward highly active users. To accurately reflect the attitude of an average American Twitter user, we report a two-stage daily average: we first calculate the daily average sentiment toward China for every individual user and then average over all individuals to obtain the daily attitude of the population.

\bibliography{references.bib}

\clearpage
\newpage
\beginsupplement

\section*{Supplementary Information}
\setcounter{page}{1}







\begin{table}[ht]
\begin{center}
\begin{tabular}{ c | c }
    Tweet text & Sentiment score \\ \hline
    Get Well Soon \#Wuhan Hope for the best. Good Luck China govt. & Most favorable (2) \\ \hline
    Iran is spreading coronavirus faster than China did. & \\
    Dunno if its incompetence or deliberate. & \\
    Now I appreciate how China managed to contain the virus & Somewhat favorable (1) \\
    and prevent it from spreading & \\ \hline
    China encourages people to return to work... requiring & \\
    AI smartphone software that dictates whether they should & Neutral (0) \\
    be quarantined & \\ \hline
    It’s fucked up man. I agree this is the saddest thing I’ve  & \\
    seen in a while but my point is China is so desperate that  & Somewhat unfavorable (-1) \\
    they are trying to stop the disease anyway they can think of.  & \\
    They are also mass killing pigs which is really sad  & \\ \hline
    Made by a company that runs sweatshops in China & Most unfavorable (-2) \\
    \hline
\end{tabular}
\end{center}
\caption{\textbf{Examples of text sentiment on social media.} \normalfont{Sentiment values range from ``most favorable'' to ``most unfavorable''.}}
\label{tab:sentiment-sample}
\end{table}

\begin{table}[ht]
\begin{tabular}{l|l|l|l|l}
 & Observation & Observation &  & Observation \\
 & in week 1 & in week 2 &  & in week t \\ \hline
Those who are & \multirow{2}{*}{$y_1(t=1)$} & \multirow{2}{*}{$y_1(t=2)$} & \multirow{2}{*}{$\cdots$} & \multirow{2}{*}{$y_1(t=t)$} \\
treated in week 1 &  &  &  &  \\ \hline
Those who are & \multirow{2}{*}{$y_2(t=1)$} & \multirow{2}{*}{$y_2(t=2)$} & \multirow{2}{*}{$\cdots$} & \multirow{2}{*}{$y_2(t=t)$} \\
treated in week 2 &  &  &  &  \\ \hline
$\cdots$ & $\cdots$ & $\cdots$ & $\cdots$ & $\cdots$ \\ \hline
Those who are & \multirow{2}{*}{$y_d(t=1)$} & \multirow{2}{*}{$y_d(t=2)$} & \multirow{2}{*}{$\cdots$} & \multirow{2}{*}{$y_d(t=t)$} \\
in week d &  &  &  &  \\
\end{tabular}
\caption{\textbf{Notation for regression discontinuity design and difference in difference strategies}. \normalfont{$y_d(t)$ measures the macro average attitude in week $t$ among individuals who are treated in week $d$. Source: Table 4 in~\citet{brand200711}.}}
\label{tab:rdd-did-notation}
\end{table}

\end{document}